\documentclass[prfluids,groupedaddress,floatfix]{revtex4-2}

\usepackage{graphicx}
\usepackage{dcolumn}
\usepackage{bm}
\usepackage{natbib}
\usepackage{amssymb}
\usepackage[dvipsnames]{xcolor}

\usepackage[draft]{changes}
\usepackage{lineno}

\begin{document}


\title{Entrainment into particle-laden turbulent plumes}


\author{Craig D. McConnochie}
\email{craig.mcconnochie@canterbury.ac.nz}
\affiliation{Department of Civil and Natural Resources Engineering, University of Canterbury, Christchurch, New Zealand}

\author{Claudia Cenedese}
\affiliation{Physical Oceanography Department, Woods Hole Oceanographic Institution, Woods Hole, MA, USA}

\author{Jim N. McElwaine}
\affiliation{Department of Earth Sciences, University of Durham, Durham, UK}
\affiliation{Planetary Science Institute, Tucson, AZ, USA}


\date{\today}

\begin{abstract}
We use laboratory experiments to investigate the entrainment of ambient fluid into an axisymmetric turbulent plume containing dense particles with a settling velocity that is considerably smaller than the plume velocity.
We consider the effect of particle size, particle concentration, and the orientation of the plume buoyancy flux -- either in the same or opposite direction to the particle buoyancy flux.
When the plume buoyancy flux is in the opposite direction to the particle buoyancy flux, entrainment into the plume increases by up to 40\%.
The rate of entrainment increases linearly with the ratio of particle buoyancy flux to fluid buoyancy flux but does not depend on the particle size for the range of particle sizes investigated here.
In contrast, when the plume buoyancy flux is in the same direction as the particle buoyancy flux, entrainment into the plume is unaffected by the addition of particles.
The observed increase in entrainment, when the plume and particle buoyancy fluxes are in opposite directions, is consistent with inertial clustering whereby particles are ejected from regions of relatively high vorticity and accumulate in regions of relatively high strain rate within the turbulent flow field. 
Regions with high particle concentration can then experience convective instabilities that affect the entrainment of ambient fluid into the plume. 
The differing behaviour observed based on the plume buoyancy flux orientation is also consistent with the above mechanism.
Finally, based on the laboratory findings, we propose a new expression for the entrainment coefficient to take into account the effect of suspended particles of opposing buoyancy flux to the plume buoyancy flux.
\end{abstract}


\maketitle

\section{Introduction}
Turbulent multiphase flows are common in a wide variety of geophysical and engineering situations such as volcanic clouds, aerosol particles in air, bubble plumes, and fires.  
Often the particle phase is treated as a passive tracer, potentially with the addition of a settling or rise velocity.  
However, in cases where the particle and fluid phases have significantly different densities, finite Stokes number effects can lead to complex and interesting interactions including inertial clustering (for a review of turbulent multiphase flow see \cite{Balachander10}).

Our focus in this paper is on dilute suspensions of relatively dense solid particles in an axisymmetric turbulent plume.  
This focus is motivated by the numerous geophysical cases where dense sediment particles are carried by a turbulent plume, of either water or air, such as volcanos or the subglacial plumes that form around Greenland \citep{Mankoff16}.
We note a distinction between the particle-laden plumes which we focus on and the more commonly studied particle plumes \citep[e.g.][]{Bordoloi20, Lai16}.
Within a particle-laden plume, both the continuous fluid phase and the discrete particle phase contribute to the plume buoyancy flux whereas particle plumes are generally considered to have zero buoyancy flux associated with the fluid phase.

Motivated by volcanic eruptions, many studies of particle-laden plumes have focused on the resulting particle deposition.
The potential re-entrainment of particles into the rising plume has been a particular focus.
To facilitate particle deposition from a spreading intrusion, these studies have typically considered the behaviour after a plume has reached the free surface of an experimental tank \citep{Carey88,Veitch00, Cardoso01}, or that of a plume rising through a stratified ambient such that it reaches a neutral density \citep{Apsley19}.
In contrast to this earlier work, we consider the effect of particles on a rising particle-laden plume when particle re-entrainment is not important such as during the early stages of plume development or when the sinking flow is spatially separated from the rising flow \cite{McConnochie20}.

In turbulent particle-laden flows, the particle distribution tends to become non-uniform through a process that has been termed inertial clustering or preferential clustering.  
The higher inertia of the particle phase relative to the fluid phase results in particles being ejected from regions of high vorticity and accumulating in regions of high strain rate within the turbulent flow field \citep{Balachander10}.  

Inertial clustering has generally been considered to be most important when the particle Stokes number is of order~1~\citep{Salazar08}, where the particle Stokes number is defined as
\begin{equation}
{\rm St} = \tau_p/T,
\end{equation} 
$\tau_p = \rho_p d_p^2/18\mu$ is the characteristic particle response timescale, 
$\rho_p$ is the particle density, 
$d_p$ is the particle diameter, 
$\mu$ is the fluid dynamic viscosity, 
and $T$ is a characteristic timescale of the flow.

If a flow has a broad range of time scales then there will be no clustering at the smallest time scales where ${\rm St}\gg 1$, there will be the most clustering at intermediate time scales where ${\rm St}=O(1)$, and there will again be no clustering at large time scales where ${\rm St}\ll1$ \citep{Tom19,Yoshimoto07}.
In any particular flow this complete range may or may not exist.

Due to inertial clustering and several other mechanisms, the addition
of particles can modulate the turbulent properties of a flow
\citep{Balachander10}.  In turbulent plumes, one of the most important
effects of turbulence is to entrain ambient fluid into the plume.
This process is typically modelled by an entrainment coefficient
defined as $\alpha = u_e/u$ where $u_e$ is the velocity with which
ambient fluid is drawn into the plume and $u$ is the characteristic
plume velocity \citep{Morton56}.  Typically the value of the
entrainment coefficient has been found to be constant and
approximately equal to 0.12 \citep{Carazzo06}.  However, the
applicability of this value has not been tested in the case of a
particle-laden plume with clustering of particles. In this case
turbulence modulation by the particle phase could alter the
entrainment of ambient fluid into the plume.

In this paper we use measurements of the plume volume flux as a
function of distance from the source to infer the entrainment
coefficient for a variety of particle loadings.  The inferred
entrainment coefficients will be used to diagnose whether turbulence
modulation by suspended particles is significant to the entrainment of
ambient fluid into a turbulent plume.  In \S\ref{sec:Theory} we
summarize the equations regulating the fluid and particle motion. In
\S\ref{sec:Experiments} we describe the laboratory experiments that
were conducted to measure the entrainment coefficient for a variety of
particle sizes and concentrations and in \S\ref{sec:EResults} we
present the results of those experiments.  In \S\ref{sec:Discussion}
we discuss the results and the potential physical mechanisms behind
them.  Finally, in \S\ref{sec:Conclusion} we summarize the study and
highlight some open questions.

\section{Theory}
\label{sec:Theory}

\subsection{Conservation equations for a turbulent axisymmetric plume}

Turbulent axisymmetric plumes are frequently modelled by the
conservation equations for volume flux $ Q = \pi b^2 u $, specific
momentum flux $ M = \pi b^2 u^2 $, and buoyancy flux
$ B = \pi b^2 g' u$ \citep{Morton56}:
\begin{eqnarray}
\frac{dQ}{dz} &=& 2\pi b u_e, \label{eq:MortonQ}\\
\frac{dM}{dz} &=& \pi b^2 g',\\
\frac{dB}{dz} &=& 0 \label{eq:MortonB}.
\end{eqnarray}
Here $ b $ is the the top-hat plume width, $ u $ is the top-hat plume
velocity, $ g' = g(\rho_a-\rho_f)/\rho_a $ is the top-hat reduced gravity,
$ g $ is the acceleration due to gravity, $ \rho_f $ is the plume
fluid density, $ \rho_a $ is the ambient fluid density, and
$ u_e = \alpha u$ is the velocity with which ambient fluid is
entrained into the plume.
The key parameter in terms of the development of plume properties is
the entrainment coefficient, $ \alpha $, and finding a suitable value
has been the subject of a wide body of work over the past decades.
\citet{Carazzo06} tabulate a range of studies investigating the value
of the entrainment coefficient in turbulent plumes and report typical
measurements in the range of 0.12\,--\,0.13.


Despite the near ubiquity of the above model in studies of jets and
plumes, it is informative to consider an alternate model of plume
behaviour.  As described above, \citet{Morton56} considered the
conservation of volume flux, momentum flux, and buoyancy flux and used
an entrainment coefficient to model the entrainment velocity.  In
contrast \citet{Priestley55} considered the conservation of momentum
flux, buoyancy flux, and mean kinetic energy flux and used a
parameterisation of turbulent kinetic energy production.  It has been
shown that these models are somewhat equivalent but that the latter
model gives an entrainment coefficient that is explicitly a function
of the plume forcing \citep{Fox70}:\begin{equation}\label{eq:alpha_vR}
\alpha = \alpha_j + (\alpha_p - \alpha_j)\Gamma,
\end{equation}
where $ \alpha_j $ is the entrainment coefficient of a pure jet,
$ \alpha_p $ is the entrainment coefficient of a pure plume,
$ \Gamma = {\rm Ri}/{\rm Ri}_p $ is the ratio between the bulk
Richardson number of the flow and the bulk Richardson number of a pure
plume, ${\rm Ri} = bg'/w^2 = BQ^2/\theta M^{5/2}$, and $ \theta $ is a
profile coefficient of the momentum field \citep[for a detailed
description see][]{Reeuwijk15}.

Equation~(\ref{eq:alpha_vR}) separates the entrainment coefficient into
two components.  The first component is common to both jets and plumes
and is independent of buoyancy.  The second component is related to
the buoyancy forcing of a plume as described by Ri.  Recent analysis
has directly considered the energetics of a plume in order to gain a
better understanding of these two components \citep{Reeuwijk15}.  It
was found that the first component, common to both jets and plumes, is
associated with turbulent kinetic energy production while the second
component is due to the buoyancy of the flow.  A third process
which can lead to further entrainment is associated with streamwise
changes in the profiles of velocity and buoyancy.  This third
component is zero if the flow is self-similar but can be important for
unsteady flows \citep{Scase12} or for steady flows close to the source \citep{Kaminski05,Carazzo06}.  The key
implication of this energetics based understanding of the entrainment
coefficient is that it elucidates three processes whereby the
entrainment coefficient could be altered by suspended particles:
changing the turbulent kinetic production, changing the mean flow
buoyancy structure, or causing streamwise changes in the velocity and
buoyancy profiles.



In addition to the plume processes mentioned above, the flow geometry
can also have a significant impact on the entrainment coefficient.
For example, measurements of the entrainment coefficient in a
two-dimensional line plume have given values of approximately 0.22
\citep{Ellison59} while line plumes adjacent to walls have
significantly reduced entrainment coefficients of approximately 0.076
\citep{Sangras99,Parker20}.  Despite the importance of the entrainment
coefficient to plume models and the range of conditions that have been
shown to alter its value, we are not aware of any attempt to
investigate the effect of suspended particles on the entrainment
coefficient of a turbulent plume.

\subsection{Inertial effects on particle motion}

We will consider the motion of a particle in a Boussinesq, Newtonian fluid at a low particle Reynolds number
\begin{equation}\label{Re_p} {\rm Re}_p = \frac{d_p v_s}{\nu},
\end{equation}
where $ v_s $ is the particle settling velocity, $ \nu =\mu/\rho$ is the kinematic viscosity of the fluid, and $\rho$ is the plume fluid density.  
The velocity of such particles is governed by the Basset-Boussinesq-Oseen equation
\begin{widetext}
\begin{equation}
\dot{\mathbf{v}}
=
\mathbf{g} 
+ \frac{1}{\rm{St}} \frac{\mathbf{u} -\mathbf{v}}{T}
+\frac{1}{2}\beta\left(\dot{\mathbf{u}}-\dot{\mathbf{v}}\right) 
-\beta\frac{\nabla p}{\rho_f} 
+3\,\sqrt{\frac{\beta}{2\pi {\rm St}}}\int^t
\frac{\dot{\mathbf{u}}-\dot{\mathbf{v}}}{\sqrt{T}\sqrt{t-t'}}\,dt' \label{eq:BBO}.
\end{equation}
\end{widetext}
In Eq.~(\ref{eq:BBO}), $ \mathbf{u} $ and $ \mathbf{v} $ are the fluid and particle velocity vectors respectively,
$ \dot{\mathbf{u}} = \frac{\partial \mathbf{u}}{\partial t} +
\mathbf{u}\cdot\nabla \mathbf{u}$,
$ \dot{\mathbf{v}} = \frac{\partial \mathbf{v}}{\partial t}$,
$ \mathbf{g} $ is the gravity vector, $ T $ is a characteristic timescale of the flow, $ \beta = \rho_f/\rho_p $ is the ratio between the fluid and particle densities, and $ \nabla p $ is the pressure gradient. 
The terms on the right hand side of Eq.~(\ref{eq:BBO}) are, in order, gravity, drag, added mass, the Froude-Krylov force, and the Basset force.
We note that since St and $T$ are always multiplied by one another, Eq.~(\ref{eq:BBO}) can be rewritten in terms of the particle response timescale, $\tau_p$, rather than St and $T$.
Therefore, the choice of a fluid timescale is unimportant in Eq.~(\ref{eq:BBO}) provided the same timescale is used as in the definition of the Stokes number.

Considering the physical situation that we are interested in --- particles that are carried within a turbulent plume --- physical scales can be defined in order to non-dimensionalise Eq.~(\ref{eq:BBO}).  
The plume velocity $ U $ and diameter $ D $ at the source are used as the velocity and length scale while the plume fluid density $ \rho_f $ is used as the density scale.  
We note that in order for $ U $ and $ D $ to be finite we are not considering a point source plume but rather a plume that starts with an initial volume, momentum, and buoyancy flux.
$ U $, $ D $, and $ \rho_f $ can be combined to form a time scale ($ T = D/U $) and a pressure scale ($ P = U^2\rho_f $).  
Using these scales, Eq.~(\ref{eq:BBO}) can be rewritten in non-dimensional form as
\begin{widetext}
\begin{equation}\label{eq:BBO_ND}
\dot{\tilde{\mathbf{v}}}
= \tilde{ \mathbf{g}} (1-\beta)
+ \frac{1}{\rm St} (\tilde{\mathbf{u}} -\tilde{\mathbf{v}})
+\frac{1}{2}\beta\left(\dot {\tilde{\mathbf{u}}} - \dot {\tilde {\mathbf{v}}}\right)
-\beta\nabla \tilde{p}
+\frac{3}{4}\sqrt{\frac{\beta}{2\pi{\rm St}}}\int^{\tilde{t} }
\frac{\dot{\tilde{\mathbf{u}}}-\dot{\tilde{\mathbf{v}}}}{\sqrt{\tilde{t}-\tilde{t}'}}\,d\tilde{t}',
\end{equation}
\end{widetext}
where $ \tilde{\cdot} $ refers to a non-dimensional quantity.  
For the remainder of this section all variables will be non-dimensional and we will not retain the $ \tilde{\cdot} $ notation.  
For small values of St, and assuming the decay of any initial transient velocities, a power series solution to Eq.~(\ref{eq:BBO_ND}) can be written as
\begin{equation}\label{BBO_PS}
\mathbf{v} 
= \mathbf{u} + {\rm St} \left[\mathbf{g} (1-\beta) - \dot{\mathbf{u}}\right] 
- \beta {\rm St} \nabla p + O\left({\rm St}^{3/2}\right).
\end{equation}
At this order, neither the Basset force term nor the added mass term are present.

Assuming that the fluid is incompressible such that $ \nabla \cdot \mathbf{u}=0 $, the divergence of the particle velocity, to $ O\left({\rm St}^{3/2}\right) $, is
\begin{equation}\label{eq:Particle_divergence1}
\nabla \cdot \mathbf{v}
= -{\rm St} \left(q + \beta \nabla^2 p\right),
\end{equation}
where $ q = |{S}|^2 - |\Omega|^2$ is the Okubo-Weiss parameter,
$ S = \left(L+L^T\right)/2$ is the strain rate tensor,
$ \Omega = \left(L-L^T\right)/2$ is the rotation tensor, and
$ L = \nabla \mathbf{u} $.  
Taking the divergence of the Navier-Stokes equations leads to $\nabla^2 p = -q$ which allows Eq.~(\ref{eq:Particle_divergence1}) to be rewritten as
\begin{equation}\label{eq:Particle_divergence}
\nabla \cdot \mathbf{v} = -{\rm St} \,q \left(1-\beta\right).
\end{equation}
Equation~(\ref{eq:Particle_divergence}) indicates that particles that are heavier than the fluid ($\beta<1$) will concentrate in regions where the strain rate
is large compared to the vorticity (where $|S|^2>|\Omega|^2$), since in these regions $ q>0 $ and $ \nabla \cdot \mathbf{v} <0 $.  
This is the underlying justification for inertial clustering which was introduced earlier as a mechanism by which particles can modulate turbulence \citep{Balachander10}.
Inertial clustering occurs as the higher inertia of the particle phase relative to the fluid phase causes particles to be ejected from regions of relatively high vorticty ($q<0$) and to accumulate in regions of relatively high strain rate ($q>0$) within the turbulent field. 

The plume fluid equations, Eqs.~(\ref{eq:MortonQ}--\ref{eq:MortonB}), and the particle equation, Eq.~(\ref{eq:BBO}), are coupled by assuming that the particles are affected by the fluid field through Eq.~(\ref{eq:BBO}), and that the fluid field is affected by the particles only through the buoyancy term --- i.e. the fluid density, $\rho_f$, in the reduced gravity, $g'$, is replaced by the density of the fluid-particle mixture.  
Such an assumption is valid if the particle concentration is low but may cease to be true as particles cluster based on Eq.~(\ref{eq:Particle_divergence}).
This mechanism increases inhomogeneities in the buoyancy field  which forces the velocity field, increasing turbulent kinetic energy and hence the entrainment coefficient.
One manifestation of this process is convective instability which occurs with sufficiently large vertical variations in particle concentration.


\section{Methodology}
\label{sec:Experiments}

\subsection{Experimental process}

Two sets of experiments were conducted to investigate the effect of suspended particles on the entrainment coefficient of a turbulent plume.
The first set considered the case where the particle buoyancy flux acted in the opposite direction to the fluid buoyancy flux and the second set considered the case where the two buoyancy flux components acted in the same direction.
Details of all experiments are provided in table~\ref{tab:Experiments}.
Both sets of experiments measured the entrainment coefficient using a method that was similar to that used in \citet{Cenedese14}.
The method is based on the `filling box' technique described in \citet{Baines83}.  
The canonical filling box describes the propagation of a density interface through a sealed space due to a sustained turbulent plume \citep{Baines69}.
This density interface has traditionally been referred to as the first front.
\citet{Baines83} showed that if a displacement flow was added through the space, the first front will be arrested at the height where the plume volume flux is equal to the displacement volume flux.  
By iteratively changing the displacement volume flux, the plume volume flux can be measured as a function of height.  
\citet{Cenedese14} adapted this technique by first establishing a sharp density interface that was easily observable, and then using the vertical velocity of the first front in a filling box configuration to calculate the plume volume flux as a function of height. 
The variation of plume volume flux with height  gives the entrained volume flux which can be related to the entrainment coefficient.
In the experiments presented here, we have used the considerably faster method described by \citet{Cenedese14} to measure the entrainment coefficient of the plume.
 
 \begin{table}[b]
 	\caption{\label{tab:Experiments}%
 		A list of experiments and their parameters. Included are the particle concentration, $\Phi$, the symbol used for each class of experiment in later figures, the particle diameter, $d_p$, the particle Stokes number based on a bulk flow timescale and median particle diameter, St, the source volume flux, $Q_s$, the reduced gravity of the plume source fluid without particles, $g'_{\rm sal}$, the buoyancy flux associated with particles, $B_{\rm part}$,the buoyancy flux associated with salinity differences, $B_{\rm sal}$, the source buoyancy flux ratio defined in Eq.~(\ref{eq:Bratio}), $P$, the momentum jet length defined in Eq.~(\ref{eq:Lm}), $L_m$, and the source parameter, $\Gamma = (5Q^2B)/(4\alpha M^{5/2})$ \cite{Hunt01}. The Stokes number is estimated using a characteristic fluid timescale equal to $T = u/b = 0.48$\,s where $u$ and $b$ are the calculated based on Eqs.~(\ref{eq:MortonQ})--(\ref{eq:MortonB}) at a height of 20\,cm.
 	}
 	\begin{ruledtabular}
 		\begin{tabular}{ccccccccccc}
 			$\Phi$ & Symbol & $d_p$  & St & $Q_s$ & $g'_{\rm sal}$ & $B_{\rm part}$ & $B_{\rm sal}$ & $P$ & $L_m$ & $\Gamma$\\
 			\% wt. & - & $\mu m$		& - 	& $\rm cm^3\,s^{-1}$	& $\rm cm\,s^{-2}$ & $\rm cm^4\,s^{-3}$ & $\rm cm^4\,s^{-3}$ & - & cm & - \\ 
 			\colrule
 			0			& ~{\color{blue}$\bigcirc$ } & -		& -		& 3.45	& 13.4 &0 		& 46.28 		& 0 	& 3.1& 0.34\\ 
 			0			& ~{\color{blue}$\bigcirc$ } & -		& -		& 4.08	& 11.4 &0 		& 46.45 		& 0 	& 4.0 & 0.20\\ 
 			0.25   	& ~{\color{blue}$\bigcirc$ }& 38--53		&	$5.7\times 10^{-4}$	& 3.33	& 13.3 & $-4.87$ 		& 44.47 		& $-0.11$ 	& 3.2& 0.32\\ 
 			0.25   	& ~{\color{blue}$\bigcirc$ }&38--53		 &	$5.7\times 10^{-4}$	& 3.41	& 13.3 & $-4.94$ 		& 45.17 		& $-0.11$ 	& 3.3& 0.31\\ 
 			0.50   	& ~{\color{blue}$\bigcirc$ }&38--53		 &	$5.7\times 10^{-4}$	& 3.28	& 13.7 & $-9.56$ 		& 44.87 		& $-0.21$ 	& 3.3& 0.30\\ 
 			0.75   	& ~{\color{blue}$\bigcirc$ } &38--53		&	$5.7\times 10^{-4}$	& 3.51	& 13.4 & $-15.30$    & 47.12 		  & $-0.32$ 	& 3.8& 0.22\\ 
 			1.00  	& ~{\color{blue}$\bigcirc$ } &38--53		&	$5.7\times 10^{-4}$	& 3.45	& 13.6 & $-20.03$    & 46.93 		  & $-0.43$ 	& 4.1& 0.20\\ 
 			1.25   	& ~{\color{blue}$\bigcirc$ } &38--53		&	$5.7\times 10^{-4}$	& 3.45	& 13.6 & $-25.00$    & 46.83 		  & $-0.53$ 	& 4.5& 0.16\\ 
 			1.50   	& ~{\color{blue}$\bigcirc$ } &38--53		&	$5.7\times 10^{-4}$	& 3.64	& 13.6 & $-31.56$    & 49.35 		  & $-0.64$ 	& 5.4& 0.11\\ 
 			1.75   	& ~{\color{blue}$\bigcirc$ } &38--53		&	$5.7\times 10^{-4}$	& 4.17	& 22.6 & $-42.42$    & 94.25 		  & $-0.45$ 	& 3.9& 0.22\\ 
 			2.00   	& ~{\color{blue}$\bigcirc$ } &38--53		&	$5.7\times 10^{-4}$	& 4.17	& 22.8 & $-48.33$    & 95.17 		  & $-0.51$ 	& 4.1& 0.19\\ 
 			2.25	& ~{\color{blue}$\bigcirc$ } &38--53		&	$5.7\times 10^{-4}$	& 4.17	& 23.1 & $-54.33$    & 96.13 		  & $-0.57$ 	& 4.3& 0.17\\ 
 			2.25   	& ~{\color{blue}$\bigcirc$ } &38--53		&	$5.7\times 10^{-4}$	& 4.17	& 22.7 & $-54.33$    & 94.71 		  & $-0.57$ 	& 4.4& 0.17\\ 
 			0.50   	& {\color{red}$\triangle$} &53--75		&	$1.1\times 10^{-3}$	& 3.45	& 13.6 & $-10.03$    & 47.03 		  & $-0.21$ 	& 3.5& 0.27\\ 
 			0.75   	& {\color{red}$\triangle$} &53--75		&	$1.1\times 10^{-3}$	& 3.39	& 13.4 & $-14.81$    & 45.56 		  & $-0.33$ 	& 3.7& 0.24\\ 
 			1.00   	& {\color{red}$\triangle$} &53--75		&	$1.1\times 10^{-3}$	& 3.39	& 13.4 & $-19.69$    & 45.29 		  & $-0.43$ 	& 4.1& 0.20\\ 
 			1.25   	& {\color{red}$\triangle$} &53--75		&	$1.1\times 10^{-3}$	& 3.39	& 13.5 & $-24.58$    & 45.66 		  & $-0.54$ 	& 4.5& 0.16\\ 
 			0.50   	& {\color{ForestGreen}$\square$} &53--75		&	$1.1\times 10^{-3}$	& 4.08	& 14.7 & $-11.92$    & 60.04 		  & $-0.20$ 	& 3.9& 0.21\\ 
 			1.00  	& {\color{ForestGreen}$\square$} &53--75		&	$1.1\times 10^{-3}$	& 4.08	& 17.2 & $-23.80$    & 70.12 		  & $-0.34$ 	& 4.0& 0.20\\ 
 			1.25   	& {\color{ForestGreen}$\square$} &53--75		&	$1.1\times 10^{-3}$	& 4.00	& 18.6 & $-29.08$    & 74.44 		  & $-0.39$ 	& 3.9& 0.21\\ 
 			1.50   	& {\color{ForestGreen}$\square$} &53--75		&	$1.1\times 10^{-3}$	& 4.08	& 20.2 & $-35.59$    & 82.45 		  & $-0.43$	& 4.0& 0.21\\ 
 			1.75   	& {\color{ForestGreen}$\square$} &53--75		&	$1.1\times 10^{-3}$	& 4.17	& 21.5 & $-42.38$    & 89.38 		  & $-0.47$ 	& 4.1& 0.20\\ 
 			0.50   	& {\color{Mulberry}$\times$} &63--90		&	$1.6\times 10^{-3}$	& 4.08	& 14.6 & $-11.92$    & 59.51 		  & $-0.20$ 	& 3.9& 0.21\\ 
 			0.50   	& {\color{Mulberry}$\times$} &63--90		&	$1.6\times 10^{-3}$	& 4.08	& 13.4 & $-11.88$    & 54.69 		  & $-0.22$	& 4.1& 0.19\\ 
 			1.00   	& {\color{Mulberry}$\times$} &63--90		&	$1.6\times 10^{-3}$	& 4.08	& 13.4 & $-23.70 $   & 54.71 		  & $-0.43$ 	& 4.9& 0.14\\ 
 			0   	& {\color{Cyan}$\diamond$} &-	&	-	& 3.64	& 13.7 & $0 $   & 49.97 		  & $0$ 	& 3.2& 0.32\\ 
 			0   	& {\color{Cyan}$\diamond$} &-	&	-	& 3.57	& 13.5 & $0 $   & 48.29 		  & $0$ 	& 3.2& 0.32\\ 
 			0.50 	& {\color{Cyan}$\diamond$} & 38--53	& $5.7\times 10^{-4}$	& 3.57	& 10.9 & $10.39 $   & 38.85 		  & $0.27$ 	& 3.2& 0.32\\ 
 			1.00   	& {\color{Cyan}$\diamond$} & 35--53	& $5.7\times 10^{-4}$	& 3.64	& 7.9 & $21.14 $   & 28.66 		  & $0.74$ 	& 3.2& 0.31\\ 
 			1.00   	& {\color{Cyan}$\diamond$} & 35--53	& $5.7\times 10^{-4}$	& 3.57	& 7.7& $20.75 $   & 27.35 		  & $0.76$ 	& 3.2& 0.31\\ 
 			1.60   	& {\color{Cyan}$\diamond$} & 35--53	& $5.7\times 10^{-4}$	& 3.57	& 4.6 & $33.15 $   & 16.26 		  & $2.04$ 	& 3.2& 0.32\\ 
 		\end{tabular}
 	\end{ruledtabular}
 \end{table}


All experiments were conducted in a tank with a 40\,cm square cross section and a height of 70\,cm.
A projector was placed 7.0\,m behind the tank and the shadowgraph technique
was used to visualise the experiment.  This allowed both the turbulent
plume and the ambient stratification to be visualised over the course
of an experiment.

\begin{figure}
	\centering
	\includegraphics[width=0.3\textwidth]{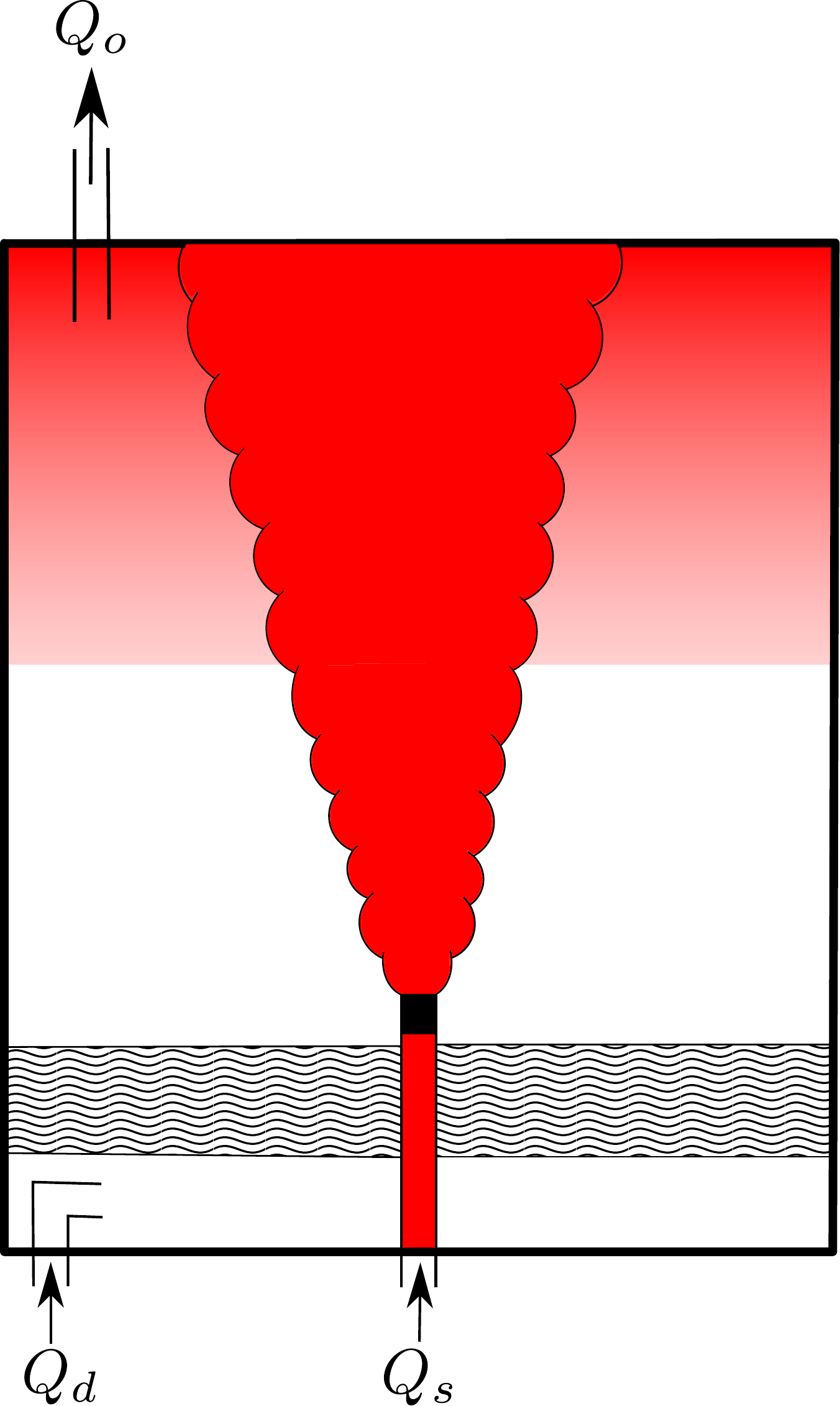}
	\caption{A schollaematic of the tank used in the experiments. $Q_s$ provided the source fluid for the plume and was released from above the foam mesh (marked by the black square). $Q_d$ provided a displacement flow that was used to arrest the first front until it was clear and sharp. $Q_o$ provided an overflow to maintain a constant volume of fluid in the tank.}
	\label{fig:schematic}
\end{figure}

A schematic of the experimental tank is shown in figure~\ref{fig:schematic}.  
The tank was fitted with two sources of fluid at the base of the tank, $Q_d$ and $Q_s$.  
One of these ($Q_s$) was used to produce the turbulent plume and was placed in the center of the tank and oriented vertically upward.  
It was designed such that the plume was turbulent upon leaving the source \citep[see][]{Cenedese14}.  
The radius of the source was $0.27\,\rm{cm}$ and the source volume flux is provided in table~\ref{tab:Experiments}.  
Selection of the source volume flux required a balance between minimising the momentum jet length, maintaining turbulence within the plume immediately from the source, and, particularly at large particle concentrations, avoiding particles becoming trapped within the tubing leading to the source.
As shown on table~\ref{tab:Experiments}, experiments were conducted with two different nominal source volume fluxes, $3.5\,{\rm cm^3\,s^{-1}}$ and $4.1\,{\rm cm^3\,s^{-1}}$. 
No dependence on the entrainment coefficient was observed, regardless of whether the reduced gravity was adjusted to maintain a constant buoyancy flux or not.

The second source ($Q_d$) was oriented horizontally and was placed underneath a layer of solid but permeable
foam.  
The second source was used to provide the displacement flow that arrests the first front \citep{Baines83}.
Over time, the first front became sharp and easily identifiable so that it's unsteady position could be tracked during the measurement stage of an experiment.
Just below the top of the tank was an overflow ($Q_o$) that allowed the volume of fluid in the tank to be kept constant while the first front was being established and an experiment was being run.

The tank was initially filled with oceanic salt water
($ \rho_a = 1.02186\pm 0.00011\,{\rm g\,cm^{-3}} $) that had been left
for at least 12 hours in a storage tank to thermally equilibrate.  The
plume source fluid was made from a mixture of oceanic salt water and fresh
water which had also been thermally equilibrated for at least 12
hours.  A small quanitity of food dye was added to the plume solution
for visualisation.  The density of both the source fluid ($\rho_s$)
and the ambient fluid ($\rho_a$) were measured with an Anton Parr
densimeter to an accuracy of $5 \times 10^{-5}\,\rm{g\,cm^{-3}}$.

After the fluid density was measured, the plume source fluid was split into two storage tanks.  
To one of these a known mass of solid glass microspheres was added to achieve the desired particle concentration.
The microspheres had a density of $2.5\,\rm{g\,cm^{-3}}$ and diameter ranges of $38$--$53$, $53$--$75$, and $63$--$90\,\rm{\mu m}$.  
These particle sizes will hereafter be signified within table~\ref{tab:Experiments} and later figures by blue circles, red triangles, and purple crosses, respectively.  
Representative settling velocities for the smallest and largest particle ranges were approximately $0.16$ and $0.46\,\rm{cm\,s^{-1}}$,
respectively, which is significantly smaller than typical plume velocities of $4$--$6\,\rm{cm\,s^{-1}}$.  
These settling velocities have been calculated based on the average of the maximum and minimum diameters for each particle size.

It is helpful to consider the buoyancy flux associated with salinity differences, $B_{\rm sal}$, and the buoyancy flux associated with the particle field, $B_{\rm part}$, separately.
A similar approach has been used to study sedimentation from particle-laden plumes in stratified ambient fluids \citep{Apsley19}.
We define a positive buoyancy flux to be in the direction that the plume travels such that $B_{\rm sal}>0$.
The total buoyancy flux of the plume is given by $B = B_{\rm sal} + B_{\rm part}$ and we also define a source buoyancy flux ratio, $P$, that gives the relative magnitude of the two buoyancy flux contributions:
\begin{equation}
P = \frac{ B_{\rm part}}{B_{\rm sal}}.
\label{eq:Bratio}
\end{equation}
Unlike previous work, we allow the value of $ P $ to be negative or positive to distinguish between the two experimental cases where buoyancy fluxes act in opposing directions or the same direction \citep{Apsley19}. 
For the majority of the experiments where the particle induced buoyancy flux is in the opposite direction to the salinity induced buoyancy flux, $ P $ will be negative and greater than -1.
The two buoyancy flux components, $B_{\rm sal}$ and  $B_{\rm part}$, as well as the source buoyancy flux ratio are given for each experiment in table~\ref{tab:Experiments}.

For most of the experiments, the reduced gravity ($g' = g(\rho_a - \rho_f)/\rho_a$) of the fluid used within the plume was set at approximately $13.5\,\rm{cm\,s^{-2}} $ before the addition of particles.  
In these experiments, the total buoyancy flux of the plume, given by $B_{\rm sal} + B_{\rm part}$, depended on the initial particle concentration.
To demonstrate that the buoyancy flux  of the plume did not affect the value of the entrainment coefficient, several additional experiments using the medium particle size were performed.  
In these experiments, the total buoyancy flux of the plume was kept constant by varying the density of the plume source fluid based on the particle concentration.
Within table~\ref{tab:Experiments} and later figures, this set of experiments is signified by green square markers.
These two sets of experiments allowed a direct comparison between plumes where the buoyancy flux due to salinity was kept constant and plumes where the total buoyancy flux was kept constant, thus highlighting any effect that changes in the total plume buoyancy flux could have on the measured entrainment coefficient.

At the start of an experiment, a sharp first front is formed in the manner described in \citet{Baines83}.  
Both sources of water at the base of the tank, as well as the overflow at the surface, are turned on and the first front is formed at the height where the displacement volume flux is equal to the plume volume flux.  
For this initial stage the plume source fluid that did not contain particles was used, so as to avoid increasing the particle loading immediately above the first front.  
Once the first front was sufficiently sharp and visually obvious, the particle-laden plume source fluid was supplied to the experimental tank for several minutes to ensure that particles would flow from the plume source as soon as the measurement phase of the experiment was started.

The plume source, $Q_s$, was then turned off and the displacement volume flux, $Q_d$, was reduced.  
The displacement volume flux was continued at a low flow rate to move the first front higher in the tank and hence increase the height over which the plume volume flux could be measured.  
A lower ambient fluid flow rate, $Q_d$, was used over this period to avoid mixing and ensure that the first front remained sharp.

Once the first front was approximately 40\,cm above the source, the displacement volume flux was stopped and the plume source was turned on.  
A camera placed 2.0\,m in front of the tank was used to record the experiment.  
The experiment was run until the first front was approximately 3\,cm above the source or the  first front collapsed due to particle loading.  
During a typical experiment we did not observe particles below the first front.
This was likely because the flow induced by the plume reaching the top of the tank kept the fluid above the first front well mixed.
As a result, it is expected that the concentration of particles immediately above the first front gradually increasd over time.
For experiments with high particle concentrations, the first front eventually became convectively unstable and particles rapidly fell down into the ambient fluid, advecting fluid with them (for a description of the mechanisms leading to this collapse see \cite{Carazzo12}).  
At this point the experiment was stopped as high particle concentrations throughout the tank made optical measurements difficult and led to particles being re-entrained into the plume, thus decreasing the buoyancy flux.
The collapse of the first front determined the maximum particle concentration that could be used in our experiments.

The above description relates to the first and larger set of experiments.  
As previously mentioned, a secondary set of experiments was also conducted where the experimental apparatus was essentially inverted.
Inverted experiments are signified by light blue diamonds on table~\ref{tab:Experiments} and in later figures.
The inverted experiments were conducted  to determine whether the behaviour of the plume is the same when the particle buoyancy flux is in the same or opposite direction to the plume buoyancy flux.

For these experiments $Q_s$ and $Q_d$ were moved to the top of the tank while $Q_o$ was moved to the bottom.  
The tank was initially filled with fresh water and the plume solution was still made from a mixture of fresh and oceanic salt water to obtain the required density difference.  
The source used for the inverted experiments was of the same design but different from that used in the original experiments.  
Apart from these minor details however, the experiments were identical.
For the experiments without particles, the entrainment coefficient for the original and inverted appartus was the same, within experimental uncertainty.

\subsection{Experimental analysis}

For all experiments, the plume volume flux was measured in the same manner as in \citet{Cenedese14}.  
To obtain the first front position as a function of time, the experimental video was processed by averaging vertical columns from outside of the plume and applying a light intensity threshold.  
Horizontal averaging was necessary as the first front developed wave-like oscillations due to residual flows above the first front and, as such, was not always horizontal.
This effect was particularly prominent at early stages of an experiment when the first front was close to the top of the tank.
This process identified the interface between the red plume fluid and colourless ambient fluid and provided the distance between the first front and the plume source as a function of time, $h(t)$.
The plume volume flux, $Q(z)$, was then calculated following
\citet{Cenedese14} as
\begin{equation}
Q(z=h) = -\left[A-A_p(z)\right]\frac{{\rm d} h}{{\rm d}  t} + Q_s,
\label{eq:Q}
\end{equation}
where $A$ is the area of the tank and $A_p$ is the area of the plume.
Since $A_p<0.03A$ in these experiments, we ignore the area of the plume which has a negligible effect on the results.
When analysing the experimental data, $ h(t) $ was approximated by fitting a polynomial to the measurements of the first front height, and the volume flux was found by differentiating that polynomial.

From Eqs.~(\ref{eq:MortonQ})--(\ref{eq:MortonB}) and following \citet{Cenedese14}, the volume flux of a pure plume is given by
\begin{equation}
Q  = \left(\frac{9}{10}\right)^{1/3} \frac{6}{5}\pi^{2/3}\alpha^{4/3}B^{1/3}z^{5/3},
\label{eq:QCenedese}
\end{equation}
where $B$ is the total plume buoyancy flux which is uniform with height.
Equation~(\ref{eq:QCenedese}) can be rearranged and differentiated with respect to $z$ to give an expression for the entrainment coefficient, $\alpha$:
\begin{equation}
\alpha = \frac{5}{6}\left(\frac{4}{3\pi^2}\right)^{1/4}
\left[\frac{\rm d}{{\rm d}z} \left(\frac{Q^3}{B}\right)^{1/5}\right]^{5/4}.
\label{eq:alpha}
\end{equation}
Equations~(\ref{eq:Q}) and~(\ref{eq:alpha}) provide a method for calculating the entrainment coefficient based on a measurement of the first front position, $h(t)$.

\begin{figure}
	\centering
	\includegraphics[width=0.95\textwidth]{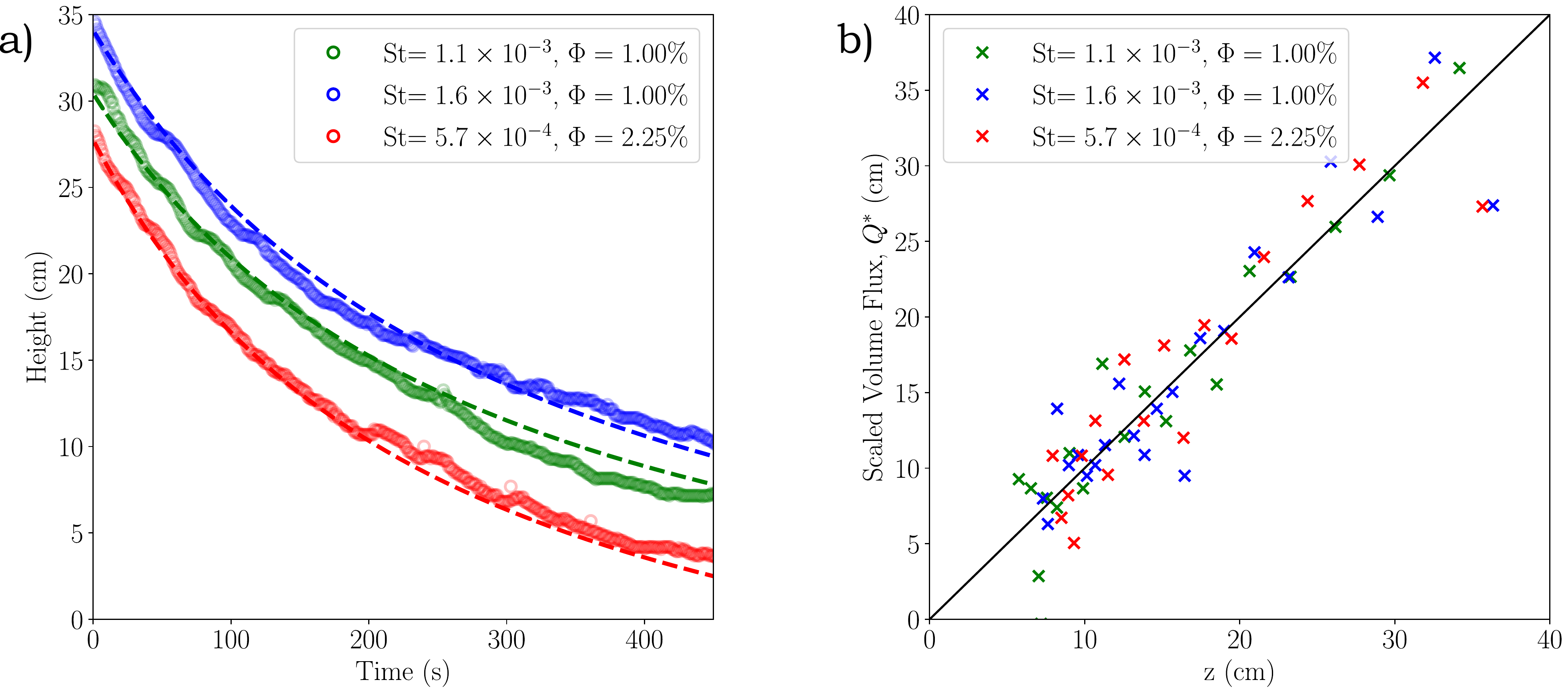}
	\caption{a) Measurements of the first front height as a function of time and that predicted based on Eq.~(\ref{eq:h(t)}) for three typical experiments.
			The profiles for ${\rm St}=1.6\times10^{-3}$ and ${\rm St}=5.7\times10^{-4}$ have been shifted vertically by a small amount to aid in clarity.
		b) Measurements of the scaled volume flux plotted against height every 30\,s for three typical experiments.  
		The scaled volume flux, $Q^*$ is given by Eq.~(\ref{eq:Q*}).  The data follow the 1:1 relationship shown by the black line with an $R^2$ value of 0.86.}
		\label{fig:Q_measurements}
\end{figure}

Implicit in our measurements of the entrainment coefficient is that the entrainment coefficient is uniform with height and that the plume follows the self-similar scaling for a pure plume as given in \citet{Morton56}, i.e. $ Q\sim \alpha^{4/3}B^{1/3}z^{5/3} $.  
This is necessary as we only measure the change in volume flux with height and hence require an \textit{a priori} model of the plume to calculate the entrainment coefficient.
To demonstrate that this self-similar scaling is valid, we have plotted two sets of intermediary result in figure~\ref{fig:Q_measurements}.
Figure~\ref{fig:Q_measurements}a shows direct measurements of the first front height as a function of time during three typical experiments.
Also shown is the first front position that would be predicted based on the self-similar scaling:
\begin{equation}
h = \left(\frac{2}{3A}\left(\frac{9}{10}\right)^{1/3} \frac{6}{5}\pi^{2/3}\alpha^{4/3}B^{1/3}t + h_0^{-2/3}\right)^{-3/2}-\frac{Q_s t}{A}
\label{eq:h(t)}
\end{equation}
where $\alpha$ is the entrainment coefficient  determined in \S\ref{sec:EResults} and $h_0$ is the height of the first front at $t=0$.
Two of the profiles in figure~\ref{fig:Q_measurements}a have been shifted vertically by a small distance to aid in visual clarity.
Figure~\ref{fig:Q_measurements}b shows direct measurements of the scaled volume flux (from Eq.~(\ref{eq:Q})) against height, every 30\,s, for the three representative experiments shown in figure~\ref{fig:Q_measurements}a. 
Here, the scaled volume flux is given by
\begin{equation}
Q^* = \left({Q} \left(\frac{9}{10}\right)^{-1/3} \left(\frac{6}{5}\right)^{-1}\pi^{-2/3}\alpha^{-4/3}B^{-1/3}\right)^{3/5}
\label{eq:Q*}
\end{equation}
 such that, provided the self-similar scaling is valid, $Q^*$ is equal to $z$.
Once again, the entrainment coefficient determined in \S\ref{sec:EResults} is used when calculating $Q^*$.
{We note that the scaled volume flux measurements shown in figure~\ref{fig:Q_measurements}b were not used to calculate the entrainment coefficient.
Figure~\ref{fig:Q_measurements}b shows the scaled volume flux directly from the experimentally measured first front heights whereas the entrainment coefficients were calculated based on the polynomial fit of $h(t)$.}

The scaled volume flux shown on figure~\ref{fig:Q_measurements}b follows the 1:1 relationship predicted by Eq.~(\ref{eq:QCenedese}), albeit with a reasonable degree of scatter ($R^2=0.86$). 
The scatter, particularly at early times (large heights), is associated with high frequency variations of the first front position and reduces over the duration of an experiment as the density difference across the first front increases.  
Despite the scatter, the data demonstrate that Eq.~(\ref{eq:alpha}) is suitable to calculate the entrainment coefficient.

\subsection{Uncertainty analysis}

\begin{figure}
	\centering
	\includegraphics[width=0.75\textwidth]{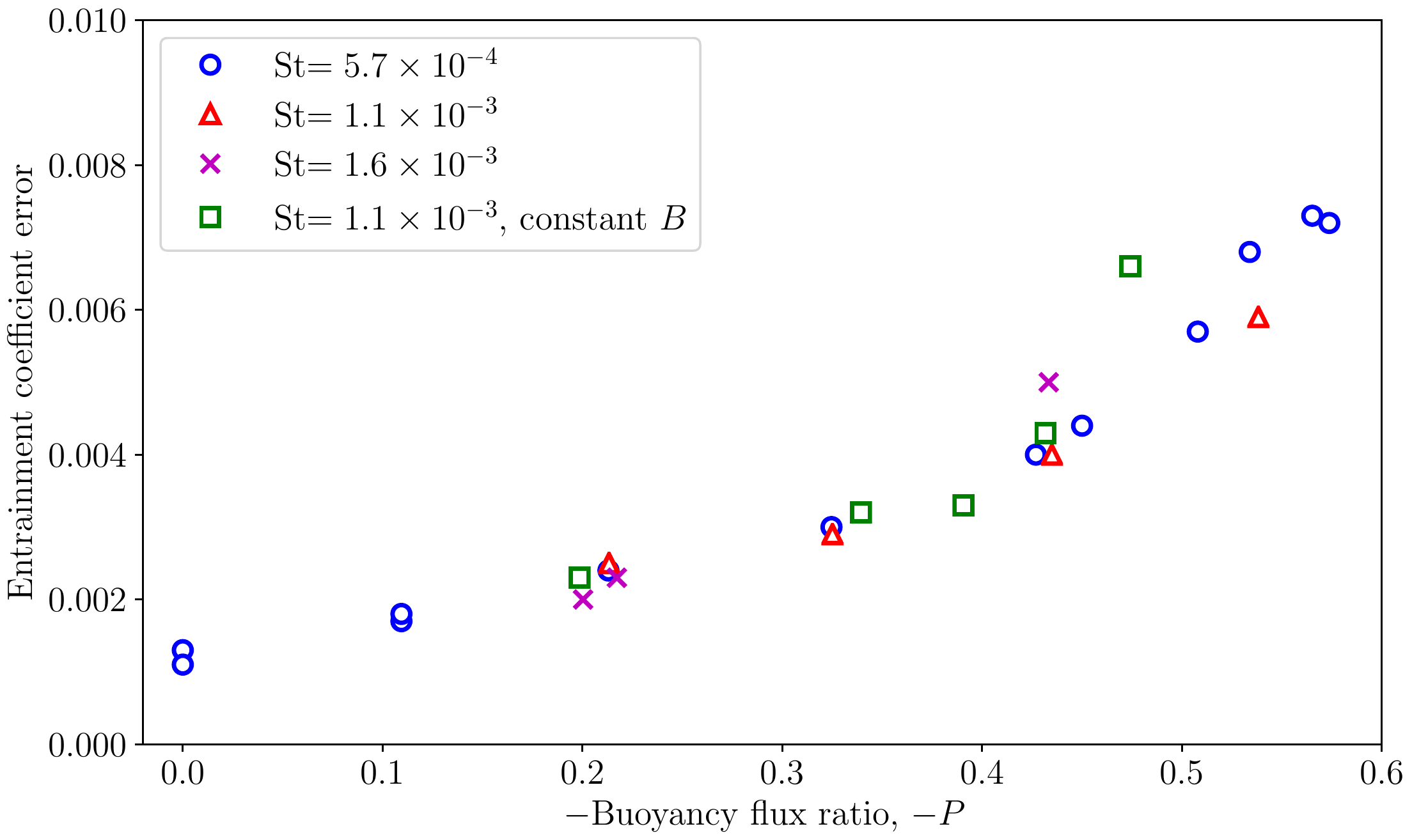}
	\caption{{Combined error on the measured entrainment coefficient as a function of the buoyancy flux ratio, $ P $.}}
	\label{fig:error}
\end{figure}

{A detailed uncertainty analysis has been conducted to assess the experimental error on the entrainment coefficient.
This analysis has included uncertainties arising both from the source conditions and from the analysis process.
In this section we describe each of the sourses of experimental error independently and then assess the combined effect.

The source volume flux was measured to within an accuracy of $\pm 0.09\,{\rm cm^3\,s^{-1}}$ which results in an error on the measured entrainment coefficient (through Eqs.~(\ref{eq:Q}) and~(\ref{eq:alpha})) of approximately $ \pm 0.0008 $.
Similarly, the fluid density difference was measured to within $ \pm 0.4\% $.
We calculate the error on the measured entrainment coefficient that results from changing the source reduced gravity by $\pm 1\%$ and find it to be approximately $\pm 0.0003$.

It is difficult to assess how accurately the source particle concentration was known.
The average particle concentration in the source solution reservoir was known to a high degree of accuracy but the actual particle concentration at the plume source was less certain.
Although the particle solution was continually stirred during the experiment, it is likely that the particle concentration at the plume source was somewhat unsteady. 
In addition, some fraction of the particles may have become trapped inside the tubing between the reservoir and the plume source.
Since it was not possible to measure the particle concentration at the source, we use a conservative estimate that the particle concentration was only known to within $ \pm 10\% $.
The uncertainty in the particle concentration affects the measured entrainment coefficient directly through Eq.~(\ref{eq:alpha}), by changing the source buoyancy flux, and also changes the buoyancy flux ratio, $ P $.
The change in the buoyancy flux results in an experimental error on the entrainment coefficient that increases with the source particle concentration between 0 and 0.0053.
The error in the buoyancy flux ratio is approximately $ \pm 10\% $.

Finally, we consider the error introduced in fitting polynomial curves to the experimental data.
Again, we note that although measurements of the volume flux are shown every 30\,s on figure~\ref{fig:Q_measurements}b, we use data of the first front height every 1\,s when calculating the entrainment coefficient.
We estimate the error in fitting both height against time (Eq.~(\ref{eq:Q})) and volume flux against height (Eq.~(\ref{eq:alpha})) by adding and subtracting two standard deviations to the fit, continuing the analysis, and recalculating the entrainment coefficient.
This introduces error on the calculated entrainment coefficient of approximately $ \pm 0.0002 $ and $ \pm 0.0003 $ for the two polynomial fits, respectively.

The exact values of each of these errors on the measured entrainment coefficient vary from experiment to experiment, most notably for the source particle concentration.
All sources of error are combined for each experiment in a way that maximises the total error on the measured entrainment coefficient.
Figure~\ref{fig:error} shows the combined error on the entrainment coefficient for the experiments with the plume buoyancy flux and particle buoyancy flux acting in opposing directions.
The same analysis was performed on experiments where the plume buoyancy flux and the particle buoyancy flux were acting in the same direction and similar results were found.
It is clear that the measured error increases as the buoyancy flux ratio becomes more negative (or the particle conentration increases) which is due primarily to the uncertainty in the source particle concentration.
The errors shown on figure~\ref{fig:error} will be used throughout later sections when presenting the experimental results.}

\section{Experimental results}
\label{sec:EResults}

\subsection{Opposing buoyancy fluxes}
\label{sec:EResults1}

\begin{figure}
	\centering
	\includegraphics[width=0.75\textwidth]{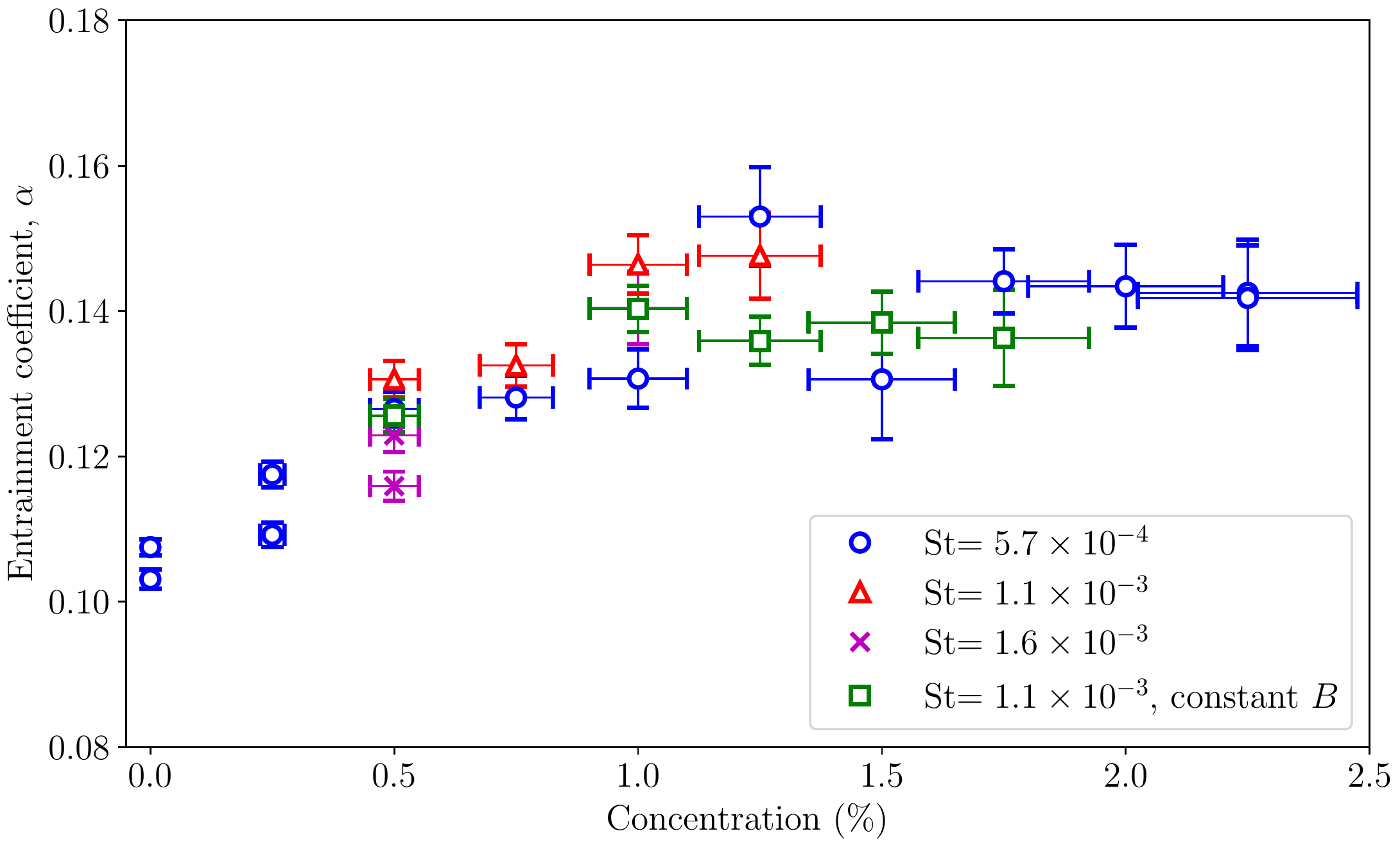}
	\caption{Measured entrainment coefficient as a function of sediment
		concentration by mass.
		{The vertical error bars show the calculated uncertainty shown in figure~\ref{fig:error} and the horizontal error bars reflect the particle concentration at the source varying by $ \pm 10\%$.}}
	\label{fig:entrainment}
\end{figure}

For the experiments where the buoyancy flux of the plume fluid and of the particles were in opposing directions, three particle sizes were used.
The results of these experiments are shown on figure~\ref{fig:entrainment} as a function of particle concentration.  
{The vertical error bars show the calculated uncertainty shown in figure~\ref{fig:error} and the horizontal error bars reflect the particle concentration at the source varying by $ \pm 10\%$.}
For low particle concentrations ($\leq 1\%$) the entrainment coefficient increases approximately linearly with increasing concentration. 
For particle concentrations above $1\%$ the entrainment coefficient appears to be approximately constant.

To ensure that source characteristics were not affecting the measured entrainment coefficients, values were calculated based on the volume flux at heights larger than the momentum jet length $ L_m $.
The momentum jet length was calculated as given in \citet{Hunt01} as
\begin{equation}
L_m = 2^{-3/2}\alpha^{-1/2}\frac{M_0^{3/4}}{B^{1/2}}
\label{eq:Lm}
\end{equation}
where $ M_0 $ is the momentum flux at the source and $ \alpha = 0.105 $ is used for all experiments.  
The value of $ \alpha=0.105 $ was chosen based on experiments with no particles.
$ L_m $ varied between approximately 3--5\,cm (10--20 source radii) across the experiments and the results were only slightly different if the entrainment coefficient was calculated over the full depth of the experiments.  
Calculating the entrainment coefficient only over heights larger than $5L_m$, based on \citet{Papanicolaou88}, had no effect on the measured entrainment coefficient.
To increase the amount of available data from which to calculate the entrainment coefficient and reduce the experimental uncertainty, all heights larger than $L_m$ were used.
In addition, there was no relationship between $ L_m $ and the particle concentration so variations in $ L_m $ can not
explain the experimental results.

All three particle sizes collapse onto a single curve in figure~\ref{fig:entrainment} demonstrating that, for the particles tested here, the entrainment coefficient does not depend on the particle size.
This is not surprising given that the settling velocity of the largest particles is only four times that of the smallest particles.
Similarly, the particle response time $\tau_p$ varied from $(1.9-7.7)\times 10^{-4}\,{\rm s}$, giving similar values of {\rm St} for the three particle size used in this study as shown in table~\ref{tab:Experiments}.
The data with constant total buoyancy flux (green squares) lie upon that with a constant buoyancy flux due to salinity differences alone.
This result suggests that the total buoyancy flux of the plume does not affect the entrainment coefficient which is consistent with studies of single phase plumes.


We note that the results shown on figure~\ref{fig:entrainment} are not consistent with an increased velocity of the first front due to the loading of particles above the front, or the sedimentation of particles through the front.  
The entrainment coefficient is determined by $ \frac{{\rm d}Q}{{\rm d}z} $ or $ \frac{\rm d}{{\rm d}z}\left(\frac{{\rm d}h}{{\rm d}t}\right) $ as given in Eqs.~(\ref{eq:Q}) and~(\ref{eq:alpha}).  
Larger values of $ \frac{\rm d}{{\rm d}z}\left(\frac{{\rm d}h}{{\rm d}t}\right) $ imply higher rates of entrainment into the plume and hence a larger entrainment coefficient.  
Throughout an experiment the first front moved towards the source and particles accumulated above the front.  
The effects of particle sedimentation through the first front are expected to be more significant later in the experiment (or at smaller heights) when the particle load within the tank is maximum.  
Thus, if sedimentation of particles through the first front was causing the front to fall more rapidly than it would due to the plume volume flux alone, $ {\rm d}h/{\rm d}t \propto Q $ would be artificially increased near the base of the tank.  
Such a change would decrease the measured value of $ {\rm d}Q/{\rm d}z $ and hence reduce the measured entrainment coefficient.  
Since the results on figure~\ref{fig:entrainment} show the opposite behaviour, particle sedimentation through the first front cannot explain the increased entrainment coefficient. 
However, it is possible that the increase in the entrainment coefficient is larger than what we measured due to particle sedimentation but, since particles were not observed to settle through the first front, this effect is expected to be small.

\begin{figure}
	\centering
	\includegraphics[width=0.75\textwidth]{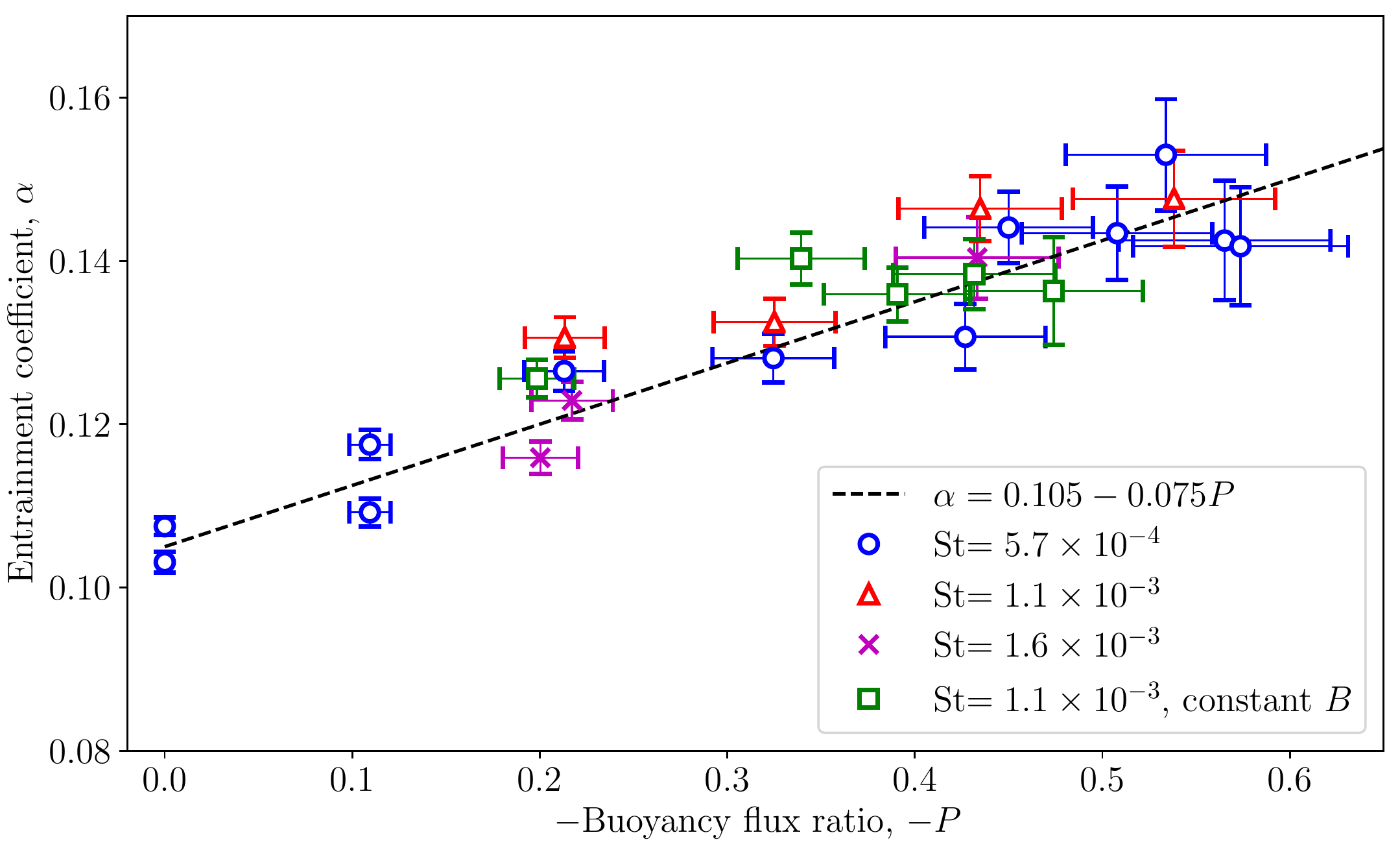}
	\caption{Measured entrainment coefficient as a function of buoyancy flux
		ratio. 
		{As in figure~\ref{fig:entrainment}, the vertical error bars show the calculated uncertainty shown in figure~\ref{fig:error} and the horizontal error bars reflect the particle concentration at the source varying by $ \pm 10\% $}.
		The empirical model (dashed line) shows a  linear fit through the data. 
		This fit will be further discussed in \S\ref{sec:Discussion}.}
	\label{fig:entrainment_Bratio}
\end{figure}


Figure~\ref{fig:entrainment_Bratio} shows the same data as plotted on figure~\ref{fig:entrainment} but instead plotted against the negative source buoyancy flux ratio, $-P$.  
While figure~\ref{fig:entrainment} suggested that the entrainment coefficient reached a constant value for particle concentrations beyond 1\% by mass, figure~\ref{fig:entrainment_Bratio} shows that the entrainment coefficient continues to increase when considered in terms of $ -P $.  
It is unclear what could cause the change in behaviour at a particle concentration of 1\% by mass.  
Therefore, the source buoyancy flux ratio appears to be the independent variable that best determines the entrainment coefficient in our experiments.

An important change in the flow regime will occur at $-P=1$, when the buoyancy flux of the particle field exactly offsets the buoyancy flux of the salinity field.
In this case, the flow would be expected to behave as a jet rather than a plume as the total buoyancy flux will be zero.
The linear model shown on figure~\ref{fig:entrainment_Bratio} would predict an entrainment coefficient of $ \alpha = 0.18 $ at $P=-1$, but the model is not expected to be valid in this regime.
The entrainment coefficient was only measured at heights greater than the momentum jet length, $L_m$.
When $-P=1$, the momentum jet length becomes infinite and the predicted entrainment coefficient would never be valid.
Whether an increase in entrainment coefficient would be observed for a particle-laden jet when compared to a single-phase jet is an important question but outside the scope of this study.

\subsection{Complementary buoyancy fluxes}
\begin{figure}
	\centering
	\includegraphics[width=0.75\textwidth]{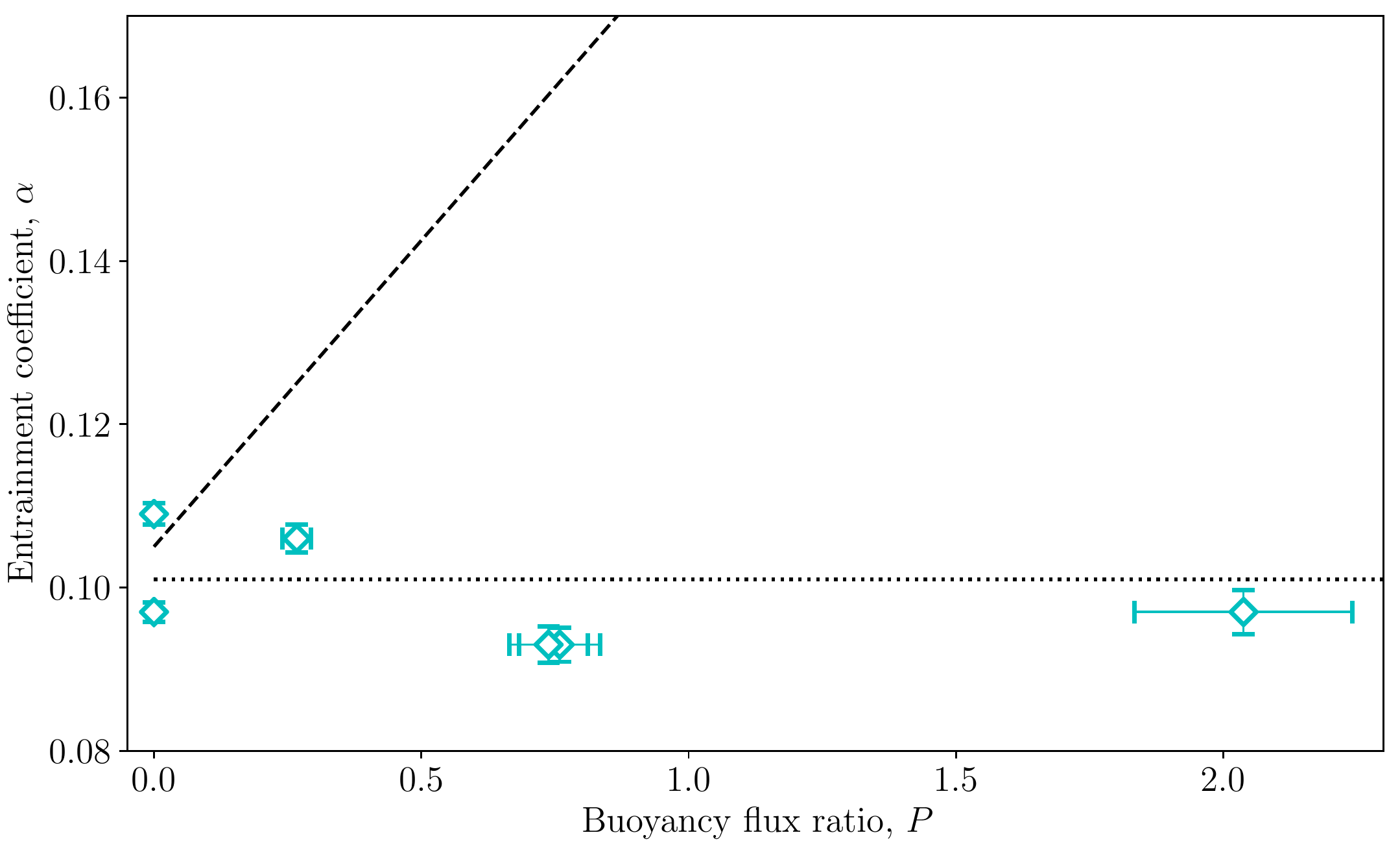}
	\caption{Measured entrainment coefficient as a function of buoyancy flux
		ratio for experiments where the plume fluid buoyancy flux and the particle
		buoyancy flux were in the same direction. 
		{The error bars have been calculated in the same manner as in figures~\ref{fig:entrainment} and~\ref{fig:entrainment_Bratio}.}
		 The dotted line shows the mean of
		the data, $\alpha = 0.101$ and the dashed line shows the same fit as was shown on figure~\ref{fig:entrainment_Bratio}. The vertical scale is the same as that used	in figure~\ref{fig:entrainment_Bratio}.}
	\label{fig:entrainment2}
\end{figure}

A second set of experiments was conducted where the buoyancy flux of the particles was in the same direction as that of the plume fluid.  
For these experiments only the smallest particle size was used and the total buoyancy flux at the source was kept constant.  The smallest particle size was used as the particles were less likely to be trapped in the piping between the fluid reservoir and the source so the experiments could be performed more reliably.  
The results of these experiments are shown on figure~\ref{fig:entrainment2}, plotted against $ P $ as that is expected to be more dynamically important than the particle concentration.

Unlike the experiments described in \S\ref{sec:EResults1}, the
measured entrainment coefficient was independent of the source buoyancy flux ratio
when the particle buoyancy flux was in the same direction as the plume
fluid buoyancy flux.  The black dashed line on figure~\ref{fig:entrainment2}
shows the mean entrainment coefficient of these experiments of 0.101
and the error bars are the same as in \S\ref{sec:EResults1}.
This value of the entrainment coefficient is slightly lower, but similar to the value $\alpha=0.105$ found for the experiments without particles in the opposing buoyancy fluxes configuration described in \S\ref{sec:EResults1}.

\section{Discussion}
\label{sec:Discussion}

The dependence of the entrainment coefficient on the particle concentration suggests that the particles are affecting the turbulent characteristics of the plume.  
\citet{Balachander10} note five different mechanisms by which particles can modulate turbulence.  
First, inertial clustering, as described in the introduction, can lead to convective instabilities in regions of high particle concentration.
Second, turbulence can be enhanced due to vortex shedding from particle wakes.  
Third, the increased inertia of the flow due to the addition of dense particles can reduce the turbulent intensity.
Fourth, the effective viscosity of the multiphase flow can be higher than that of the fluid which will reduce the turbulence. 
Finally, particle drag can increase the turbulent dissipation rate, decreasing the level of turbulent kinetic energy.  
The final three of these mechanisms will reduce the turbulent kinetic energy in the plume so are very unlikely to lead to increased entrainment.

Of the two mechanisms for turbulence enhancement, vortex shedding in
particle wakes can be discounted as a viable mechanism to increase
entrainment for the following reason.  The eddy-viscosity and
eddy-diffusivity both scale as $u_0(z)b(z)$ in a plume \citep{Morton56},
where $b(z)$ is the Gaussian width and $u_0(z)$ the Gaussian maximum velocity at height $z$. 
This can be seen from dimensional analysis since these are the only local quantities from the velocity field that can be combined to give dimensions of $L^{2}T^{-1}$.  
In addition, for an axisymmetric plume one can show consistency with an
eddy diffusivity that is the product of the velocity profile and the
plume radius. Assuming a Boussinesq flow, self-similarity and Gaussian profiles, the following axisymmetric fields for
density, vertical and radial velocity can be defined which exactly satisfy incompressibility:
\begin{eqnarray}
\rho &=& g'(z) f(s)^2,\\
u_z &=& u_0(z) f(s), \label{eq:u_z}\\
u_r &=&  \frac{r}{z}u_0(z) f(s)  - \frac{1}{4r}[1-f(s)]\frac{\partial}{\partial z}\left[u_0(z)b(z)^2\right] \label{eq:u_r},      
\end{eqnarray}
where $s=r/b$, $b=(6/5)\alpha z$ and $f(s)=\exp\left(-2s^2\right)$ . 
Equation~(\ref{eq:u_r}) is found directly from Eq.~(\ref{eq:u_z}) to satisfy incompressibility.
The choice of $f(s)^2$ for $\rho$ means that all
terms in the vertical momentum have the same exponential factor.  The
zeroth horizontal moment of the buoyancy equation requires the
conservation of buoyancy as usual with $g'(z)\propto z^{-5/3}$ and
$u_0(z)\propto z^{-1/3}$. The first horizontal moments are satisfied by
symmetry and the second horizontal moments are satisfied if the eddy
diffusivity is taken as
\begin{equation}
\frac{\kappa}{bu_z} = \frac{3}{16}\alpha +O(\alpha^3).
\end{equation}
The same can also be done with the momentum equations, if the $\alpha$
expansion is performed before the integration to $r=\infty$, this
gives an eddy viscosity
\begin{eqnarray}
\frac{\nu}{bu_z}  =  \frac{3}{4} \alpha +O(\alpha^3).
\end{eqnarray}

For a particle sedimenting at its settling velocity $v_s$ the only
length scale is the particle diameter $d$. Thus the induced eddy
viscosity/diffusivity from the turbulent wake must scale with $dv_s$.
Since the particle diameter is small compared to the plume diameter
and the settling velocity is small compared to the plume velocity,
this means that the contribution to mixing and entrainment from the
vortex shedding in particle wakes will be small. However, if particle
clustering occurs the cluster can behave like a larger particle thus
giving rise to increased eddy viscosity/diffusivity, which scales with
the product of cluster size and cluster velocity.

The observed increased entrainment coefficient is therefore most
likely explained by inertial clustering leading to convective
instabilities. Inertial clustering has been observed to have an
increased importance with increased particle concentrations
\citep{Monchaux17} which is consistent with our experiments.  In
addition, convective instabilities near the edges of the plume could
directly enhance the exchange of ambient fluid into the plume and lead
to a higher value of the entrainment coefficient.

\begin{figure}
	\centering
	\includegraphics[width=0.6\textwidth]{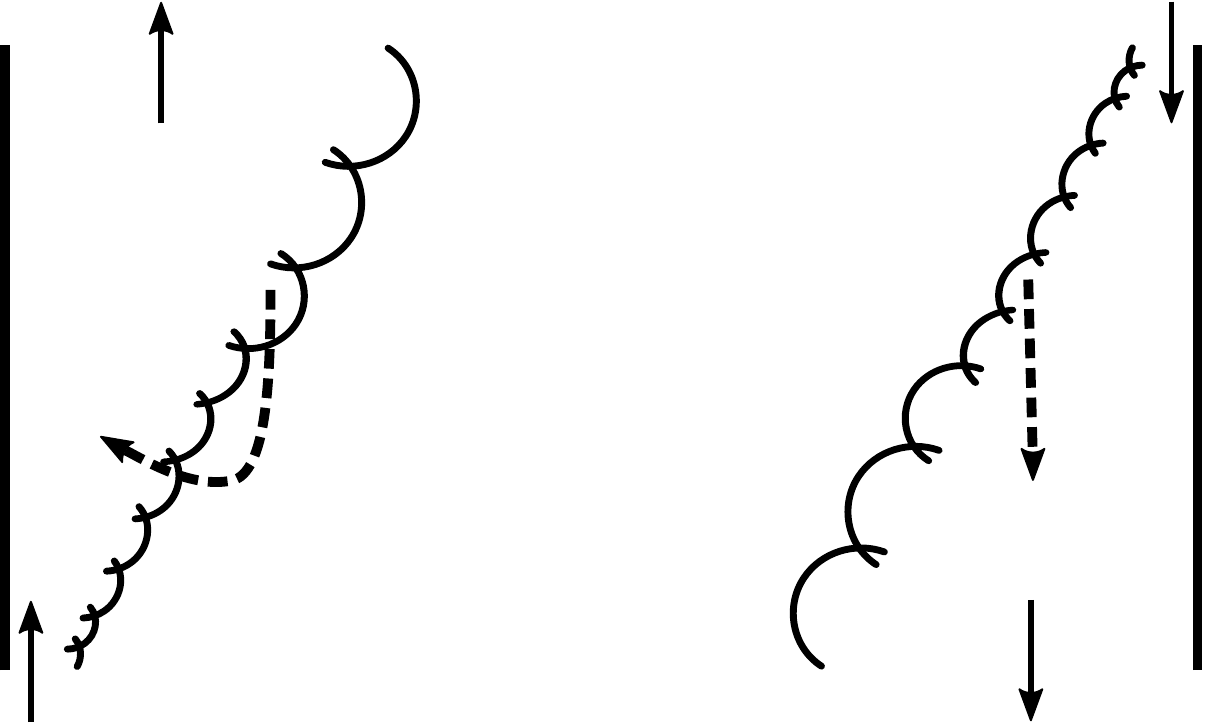}
	\caption{Schematic showing the difference between the two sets of experiments described in \S\ref{sec:EResults}.  
		{Solid arrows represent the fluid paths and dashed arrows represent the path of a dense particle cluster.}
		On the left, the case where the particle and plume buoyancy fluxes are in opposing directions is shown while on the right, the case where the	particle and plume buoyancy fluxes are in the same directions is shown.
		On the left, the convecting particles mix with ambient fluid before being reentrained back into the plume whereas on the right	the particles fall through the plume and do not encounter ambient	fluid.}
	\label{fig:entrainment_schematic}
\end{figure}

Convective instabilities arising at the edge of the plume from inertial clustering can also explain the asymmetry between the two sets of experiments described in \S\ref{sec:EResults}.  
The difference between the two seemingly comparable situations is demonstrated schematically on figure~\ref{fig:entrainment_schematic}.  
On the left of figure~\ref{fig:entrainment_schematic}, the case where the particle buoyancy flux and plume fluid buoyancy flux are in opposing directions is shown while on the right, the case where the particle buoyancy flux and plume fluid buoyancy flux are in the same direction is shown.  
For each case, we consider a small region of fluid that has reached a sufficiently high particle concentration that the local particle buoyancy exceeds the fluid buoyancy and determines the direction of the bulk density of the local region.  
The dashed lines show the path that such a dense region might be expected to follow.

In the case where the particle buoyancy flux is opposing the plume fluid buoyancy  flux (and hence the plume velocity), dense particle clusters will fall towards the plume source where the plume is narrower.  
As a result, particle clusters from the edge of the plume will fall out of the plume into ambient fluid.
These particle clusters may then be re-entrained into the plume or continue falling.
Regardless, interaction between the particle cluster and the turbulent-non turbulent interface could lead to enhanced entrainment of ambient fluid into the plume.
In contrast, when the particle buoyancy flux is in the same direction as the plume fluid buoyancy flux, particle clusters will fall away from the plume source where the plume is wider.  
As a result, particle clusters at the edge of the plume will remain within the plume, will not interact with ambient fluid, and will not be able to enhance the entrainment of ambient fluid into the plume.

As the particle concentration increases, it is expected that a smaller
particle concentration anomaly would be required for a local density
reversal to occur.  Thus, convective overturns would be able to occur
more frequently and the effect on the entrainment coefficient should
be larger.  This is consistent with the data presented on
figure~\ref{fig:entrainment_Bratio}.
We note that, during experiments, we did not observe large-scale collapse at the edge of the plume.
From this, we infer that the induced convection did not take the form of a collapsing column, as described in \citet{Carazzo12}, but instead comprised of clusters of particle-fluid mixture sedimenting sporadically from the plume over relatively short time periods.


To further investigate the changes in the entrainment coefficient with particle concentration it will be beneficial to consider the turbulent kinetic energy production directly.  
It has previously been observed that the turbulent energy production by shear is similar in jets and plumes with the only difference caused by the addition of buoyancy in a plume \citep{Reeuwijk16}.  
This observation results in expressions for the entrainment coefficient of a forced plume such as that in Eq.~(\ref{eq:alpha_vR}) which
explicitly considers the role of shear and buoyancy in producing the turbulent kinetic energy that entrains ambient fluid.  
To extend Eq.~(\ref{eq:alpha_vR}) to account for the role of buoyancy reversals driven by inertial clustering, a term that considers the particle phase is needed.  
Hence, we propose the following expression for the entrainment coefficient of a plume with suspended particles of opposing buoyancy:
\begin{equation}
\alpha = \alpha_j+(\alpha_p-\alpha_j)\Gamma -  CP.
\label{eq:alpha_new}
\end{equation}
The data presented on figure~\ref{fig:entrainment_Bratio} suggests a constant value of $ C=0.075 $ for these experiments, as shown by the dashed line.  

Equation~(\ref{eq:alpha_new}) is  empirical, based on our experimental results and the form of Eq.~(\ref{eq:alpha_vR}).
Within the experimental parameter space, treating $C$ as a constant provides an accurate prediction of measured entrainment coefficients.
However, the variation of particle properties within the experiments is relatively small and some dependence of $C$ on the particle Stokes number, the density ratio $\beta$, or other parameters is possible but outside the scope of the present study.
As previously mentioned, the relationship is also unlikely to hold as $P$ approaches $-1$ and the flow becomes a particle-laden jet.
Clearly further work in the form of a broader parametric study is required to constrain both the value of $ C $ and the form of Eq.~(\ref{eq:alpha_new}) which is currently somewhat speculative.

%
%

\section{Conclusion}
\label{sec:Conclusion}

We have presented experiments that measure the volume flux, as a function of height, of an axisymmetric turbulent plume with suspended dense particles.  
Thse measurements were used to calculate an entrainment coefficient for a variety of particle concentrations, both in the configuration where the particle buoyancy flux is in the opposite direction, and in the same direction as the plume fluid buoyancy flux.

When the particle buoyancy flux opposes that of the plume fluid, the entrainment
coefficient is seen to be significantly altered by the presence of
suspended particles, increasing linearly with the source buoyancy flux ratio. 
Its value can be up to 40\% larger than that without particles for 
the largest source buoyancy flux ratio investigated in this study, i.e. $P=0.55$.
Inertial clustering has been suggested as a
process that could lead to local buoyancy reversals and to increase the
entrainment coefficient but the experiments were not able to fully
test if this mechanism is the correct explanation.

When the particle buoyancy flux and the plume fluid buoyancy flux are in the same
direction, the entrainment coefficient appears to be unaffected by the
presence of suspended particles.  The asymmetry based on the direction
of particle buoyancy flux is consistent with inertial clustering as regions
of high sediment concentration would not lead to buoyancy flux reversals if
the particle and plume fluid buoyancy flux were aligned.

Although the experimental results can be qualitatively explained
through inertial clustering and local buoyancy reversals, the
quantitative details of this process still need to be investigated.
This would require observations of either local particle concentration
or bulk density over a variety of spatial scales.  It would also be of
interest to extend the results to a wider range of particle
properties.  It is expected that the constant, $C$, in Eq.~\ref{eq:alpha_new} is itself a function of at least the particle Stokes number and this
should be tested.  Finally, it would be interesting if similar results
would be produced if the plume was seeded with bubbles instead of
dense particles.  Particle-laden flows often have similarities with
bubbly flows so similar results might be expected in bubbly plumes.

\section*{Acknowledgements}

We gratefully acknowledge technical assistance from Anders
Jensen. CM thanks the Weston Howard Jr. Scholarship for
funding. Support to CC was given by NSF project OCE-1434041 and
OCE-1658079.
\bibliography{Sediment_laden_plumes.bib}

\end{document}